\documentclass[a4paper]{article}

\PassOptionsToPackage{hyphens}{url}\usepackage{hyperref}

\usepackage[T1]{fontenc}
\usepackage{caption, subcaption}
\usepackage{amsmath,amssymb,fixmath}
\usepackage{graphicx}
\usepackage{float}
\usepackage[dvips]{epsfig}
\usepackage{color}
\usepackage{multirow}
\usepackage{makecell}
\usepackage{threeparttable}
\usepackage{setspace}

\usepackage[
backend=biber,
style=ieee,
maxbibnames=3,
maxcitenames=3,
doi=false,isbn=false,url=false,eprint=false
]{biblatex}
\defbibheading{bibliography}[\refname]{}

\usepackage{INTERSPEECH2022}

\title{Cross-Speaker Emotion Transfer for Low-Resource Text-to-Speech\\Using Non-Parallel Voice Conversion with Pitch-Shift Data Augmentation}
\name{Ryo Terashima$^1$, Ryuichi Yamamoto$^1$, Eunwoo Song$^2$, Yuma Shirahata$^1$, Hyun-Wook Yoon$^2$,\\ Jae-Min Kim$^2$, Kentaro Tachibana$^1$}
\address{$^{1}$LINE Corp.,Tokyo, Japan, \\
$^{2}$NAVER Corp., Seongnam, Korea}
\begin{document}
\maketitle
\fontsize{8.5}{10.0}\selectfont

\begin{abstract}
Data augmentation via voice conversion (VC) has been successfully applied to low-resource expressive text-to-speech (TTS) when only neutral data for the target speaker are available.
Although the quality of VC is crucial for this approach, it is challenging to learn a stable VC model because the amount of data is limited in low-resource scenarios, and highly expressive speech has large acoustic variety.
To address this issue, we propose a novel data augmentation method that combines pitch-shifting and VC techniques.
Because pitch-shift data augmentation enables the coverage of a variety of pitch dynamics, it greatly stabilizes training for both VC and TTS models, even when only 1,000 utterances of the target speaker's neutral data are available.
Subjective test results showed that a FastSpeech~2-based emotional TTS system with the proposed method improved naturalness and emotional similarity compared with conventional methods.
\end{abstract}


\section{Introduction}

Deep learning approaches have been successfully applied to expressive text-to-speech (TTS).
Expressive styles of speech can be modeled using explicit labels for style attributes~\cite{zhu2019controlling,lei2021fine} or by extracting high-level latent features from input speech~\cite{wang2018style,skerry2018towards,hsu2019hierarchical,zhang2019learning}.
However, achieving competitive performance in low-resource scenarios remains a challenge.

In previous studies on low-resource TTS, researchers used transfer learning~\cite{jia2018transfer,tits2019exploring,chung2019semi}  or multi-speaker modeling~\cite{valle2020mellotron,Char2017DeepV2,byambadorj2021multi}.
Most recently, data augmentation techniques have been successfully applied in low-resource scenarios~\cite{hwang2021tts,huybrechts2021low,shah2021non}.
In particular, a cross-speaker style transfer method via voice conversion (VC) enables expressive TTS systems to be built where expressive data is only available for some existing speakers (i.e., source speaker)~\cite{ribeiro2022cross}. 
In this method, a pair of neutral speech databases of source and target speakers is used to learn a VC model. 
Then, the learned VC model is used to transfer the source model's expressive style (e.g., conversation) to the target speaker.
Finally, a TTS acoustic model is trained using the VC-augmented speech together with the recorded neutral speech. 

However, although a high-quality VC model is crucial for data augmentation approaches, it is challenging to learn a stable VC model when (1) the amount of data is limited under low-resource conditions or (2) highly expressive speech has large acoustic variety. 
Under such conditions, a lack of accurate prosody conversion is often observed because VC models tend to focus on spectral (e.g., Mel-spectrogram) conversion~\cite{kaneko2019CycleGAN-VC2}.
Although some VC models use a mean-variance normalization method for fundamental frequency ($F_o$) conversion~\cite{liu2007high}, this is not sufficient to stably generate the highly emotional voice of the target speaker.

To address the aforementioned problem, we propose a novel data augmentation method that combines pitch-shift (PS) augmentation and non-parallel VC-based augmentation. 
Our method differs from existing methods~\cite{ribeiro2022cross} in that the proposed system focuses on improving VC performance to make it suitable for converting emotional attributes, even though the target speaker's data only consist of neutral recordings. 

In detail, in the proposed method, we first apply PS-based augmentation to both the source and target speaker's neutral recordings.
As this enables the VC model to cover a variety of pitch dynamics, it substantially improves the stability of the training process.
Additionally, we also propose incorporating a short-time Fourier transform (STFT)-based $F_o$ regularization loss into the optimization criteria of the VC training process.
This also stabilizes the target speaker's $F_o$ trajectory, which is crucial for converting highly emotional speech segments.
As a result, the VC model learned by the proposed method stably transfers the source speaker's speaking style to the target speaker, and even makes it possible to build the target speaker's emotional TTS system.

We investigated the effectiveness of the proposed data augmentation approach by performing subjective evaluation tasks.
Note that PS-based augmentation and STFT $F_o$ regularization loss can be extended to any neural VC model; however, our focus is the Scyclone model~\cite{kanagaki2020scyclone} based on a cycle-consistent adversarial network (CycleGAN)~\cite{zhu2017unpaired}.
The experimental results demonstrated that our VC-augmented TTS system achieved better naturalness and emotional similarity than conventional methods when only 1,000 utterances of the target speaker's neutral data were available.

\section{Method}

\begin{figure}[!t]
   	\begin{minipage}[b]{1.0\linewidth}
        \centering
   	    \includegraphics[width=0.9\linewidth]{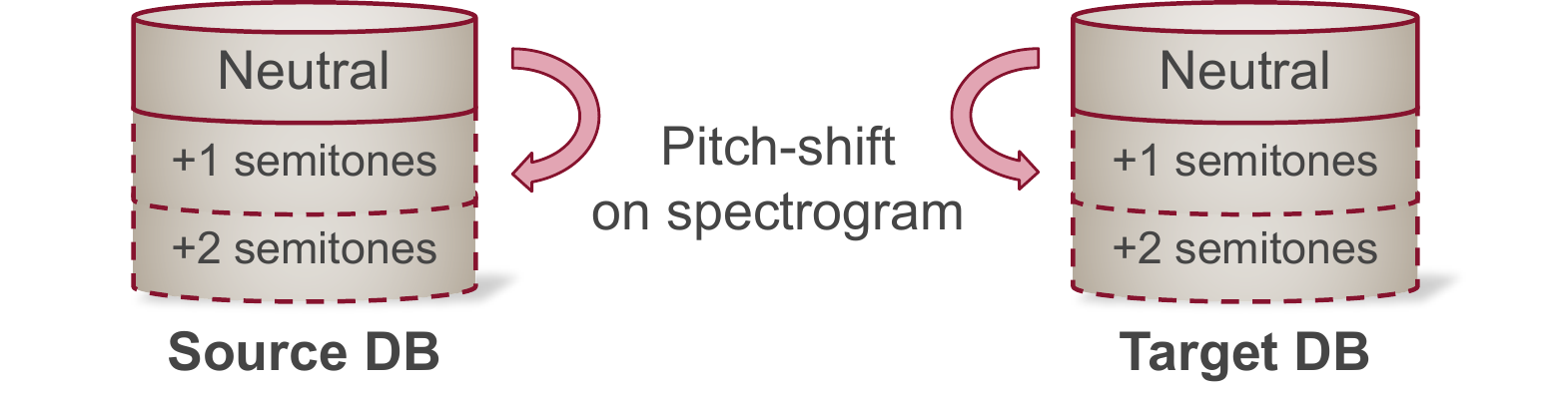}
   		\centerline{(a)}  \medskip
        \vspace{-3mm}
    \end{minipage}
   	\begin{minipage}[b]{1.0\linewidth}
        \centering
   	    \includegraphics[width=0.9\linewidth]{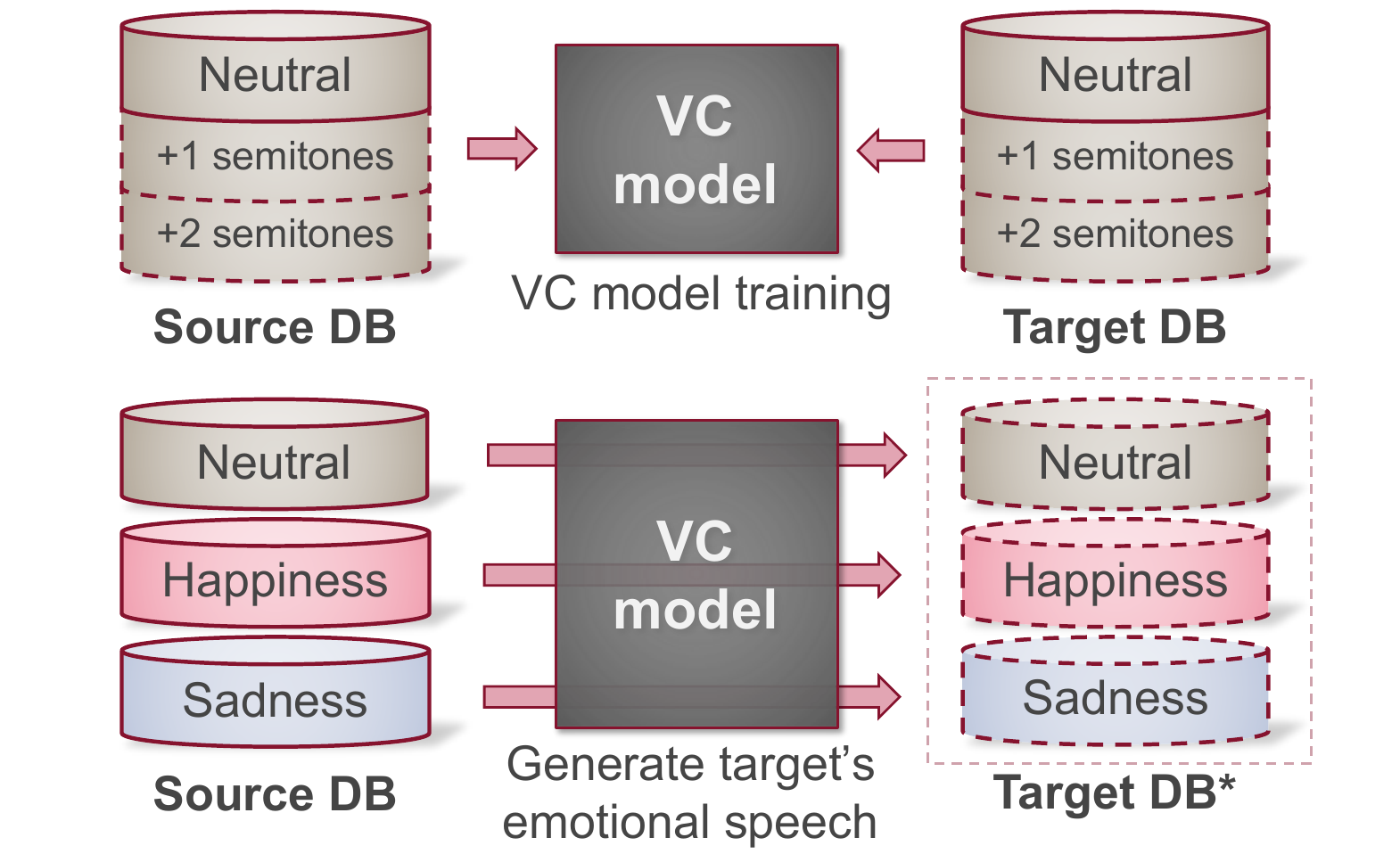}
   		\centerline{(b)}  \medskip
        \vspace{-3mm}
   	\end{minipage}
   	\begin{minipage}[b]{1.0\linewidth}
        \centering
   	    \includegraphics[width=0.9\linewidth]{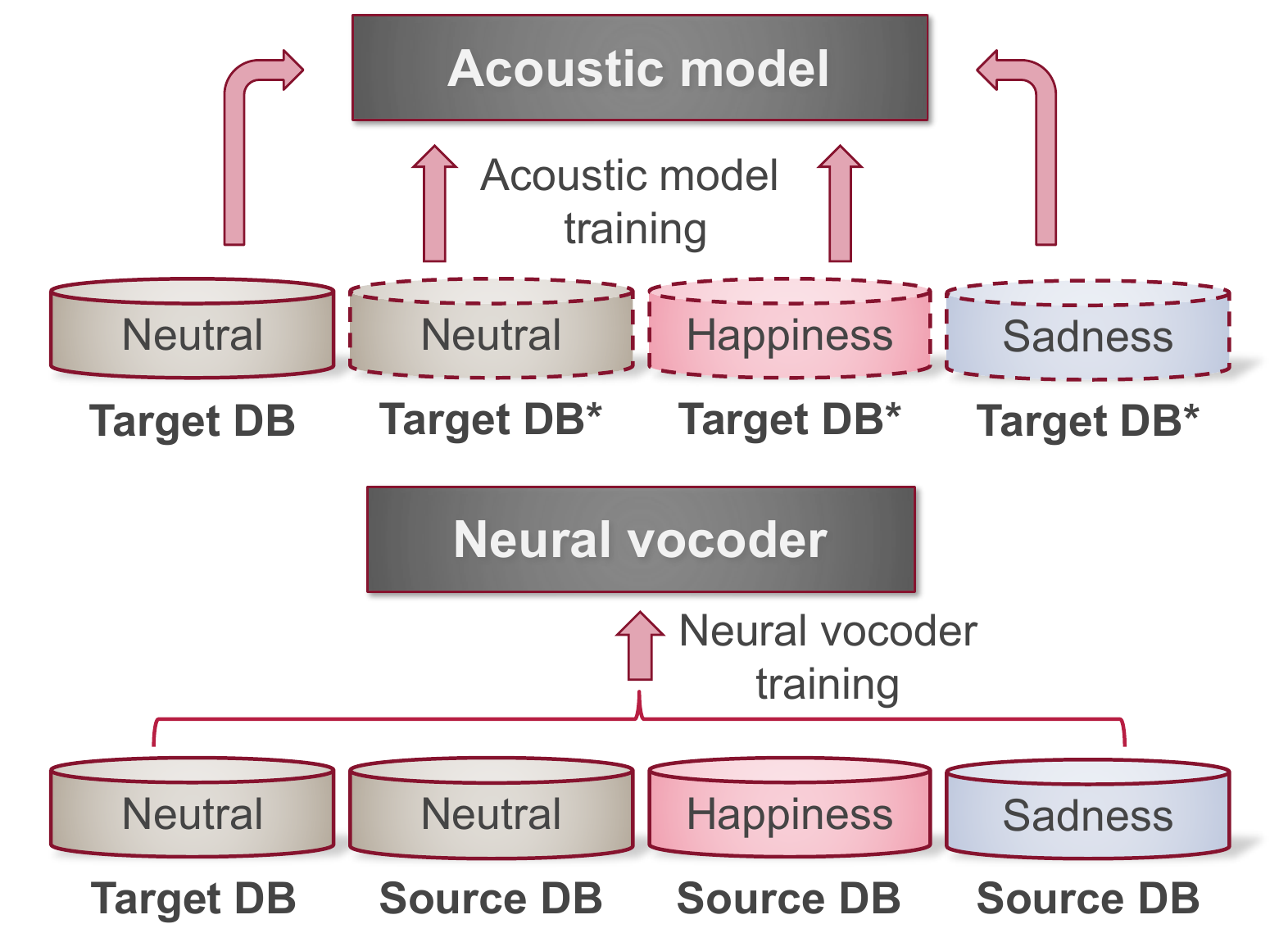}
   		\centerline{(c)}  \medskip
        \vspace{-3mm}
   	\end{minipage}
    \vspace{-8mm}
    \caption{
    Overview of the proposed method for building emotional TTS. We used data augmentation for training both the VC and TTS model: (a) PS-based data augmentation, (b) VC-based data augmentation, and (c) TTS model.
    }
    \label{fig:overview}
    \vspace{-2mm}
\end{figure}

Figure~\ref{fig:overview} shows an overview of our proposed method. 
In this study, we investigate three speaking styles: \textit{neutral}, \textit{happiness}, and \textit{sadness}.
The proposed method consists of PS-based data augmentation, VC-based data augmentation and emotional TTS system.
In the following, we describe the details of each component.

\subsection{PS-based data augmentation}

\begin{figure}[!t]
   	\begin{minipage}[b]{1.0\linewidth}
        \centering
        \includegraphics[width=0.9\linewidth]{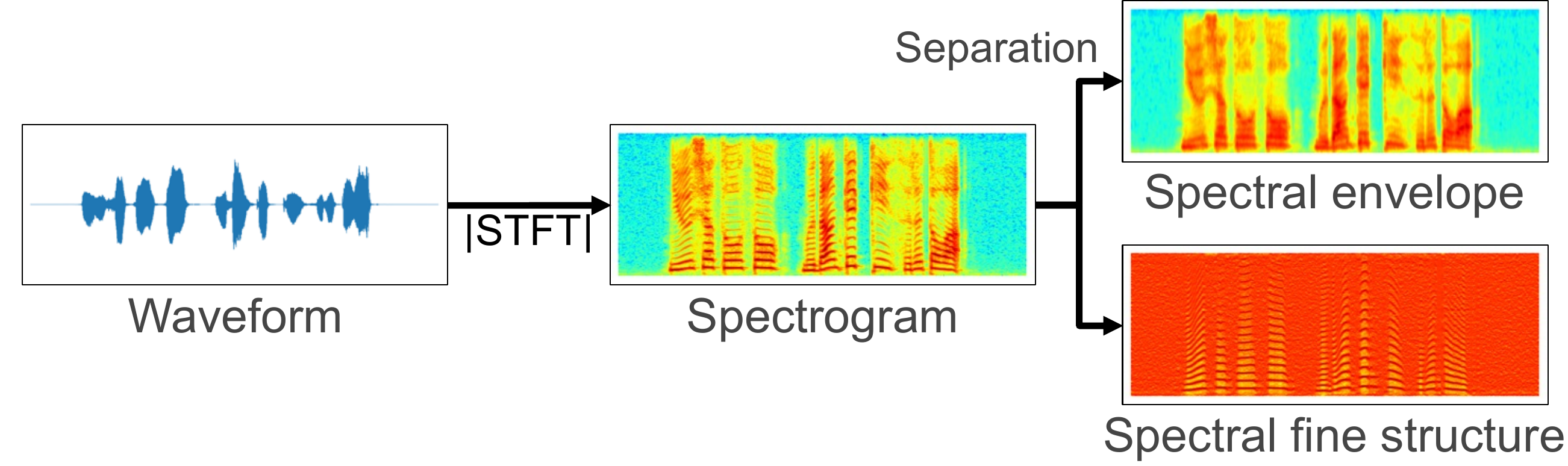}
   		\centerline{(a)}  \medskip
        \vspace{-4mm}
    \end{minipage}
   	\begin{minipage}[b]{1.0\linewidth}
        \centering
   	    \includegraphics[width=0.9\linewidth]{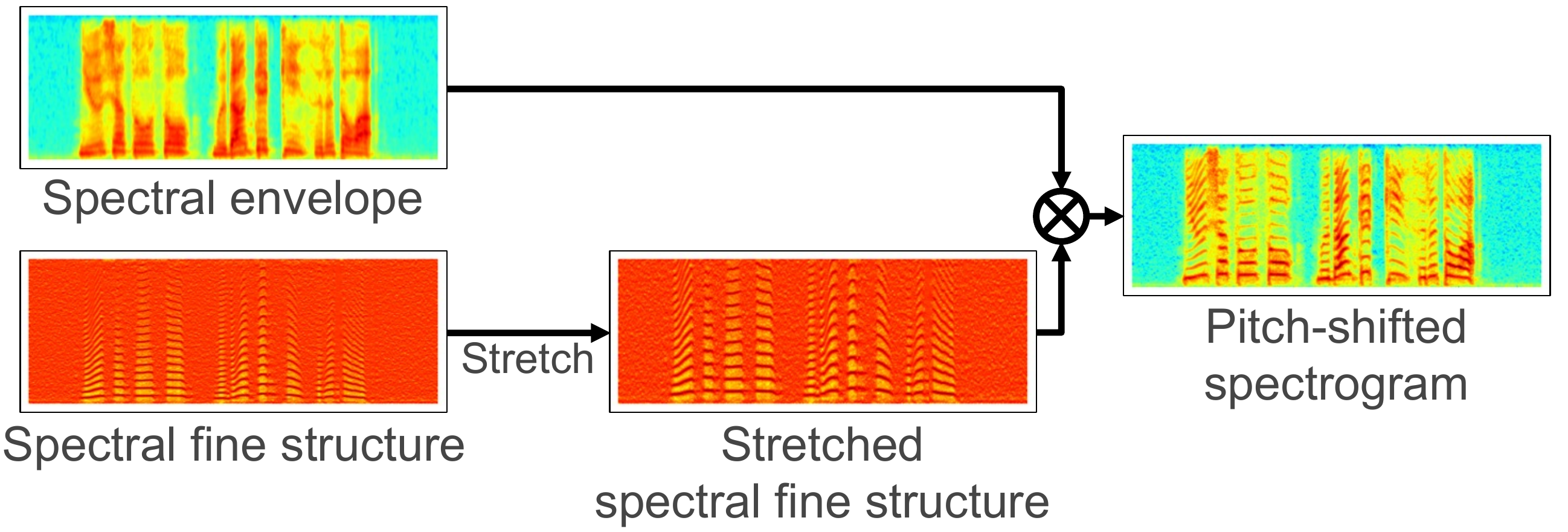}
   		\centerline{(b)}  \medskip
        \vspace{-4mm}
   	\end{minipage}
    \vspace{-8mm}
    \caption{Pitch-shift data augmentation: (a) spectrogram separation and (b) spectrogram generation.}
    \label{fig:pitch-shift}
    \vspace{-2mm}
\end{figure}

Figure~\ref{fig:pitch-shift} shows an overview of our PS-based data augmentation.
Unlike traditional PS methods such as pitch-synchronous overlap-add~\cite{charpentier1986diphone} and vocoders~\cite{Morise2016WORLDAV}, our method does not require $F_o$ estimation.
Furthermore, since it does not involve waveform synthesis, there is no need to reconstruct the phase information.
Specifically, the proposed method applies a stretching technique to the spectral fine structure to convert the pitch of the input signal.
In the separation step as shown in Figure~\ref{fig:pitch-shift}a, we first compute a speech spectrogram using STFT and then separate it into spectral envelopes and fine structures based on the lag-window method~\cite{tohkura1978spectral}.
Next, by applying a linear interpolation method, we stretch the spectral fine structure along the frequency axis.
Let $S_{t,k}$ denote the spectral fine structure for the $t$-th time index and $k$-th frequency bin.
Then, we obtain the stretched spectrum as follows~\cite{morise2018onsei}:
\begin{align}
    \hat{S}_{t,\alpha k} &= {S}_{t,k}, \label{eq:stretch} \\
    \alpha &= 2^{p/12}, \label{eq:alpha}
\end{align}%
where $\alpha$ denotes the stretching ratio determined by the semitone unit $p$.
In the generation step as shown in Figure \ref{fig:pitch-shift}b, we obtain the pitch-shifted spectrogram by multiplying the original spectral envelope and corresponding stretched spectral fine structure.

As shown in Figure \ref{fig:overview}a, we apply the proposed PS method to augment both the source and target speaker's neutral data.
In detail, we vary the semitone unit $p$ in the range [-3, 12], which results in generating data 15 times larger amount of data than the original recordings.
All the augmented datasets are used to train the VC model, which we explain further in the following section.

\subsection{Non-parallel voice conversion}
\label{ssec:vc}

\subsubsection{Model}

From the many state-of-the-art VC models, we adopt a non-parallel Scyclone model~\cite{kanagaki2020scyclone} because of its stable generation and competitive quality.
This method uses two separate modules: a CycleGAN-based spectrogram conversion model~\cite{zhu2017unpaired} and a single-Gaussian WaveRNN-based vocoder~\cite{okamoto2019real}.
However, we only use the spectrogram conversion model because VC aims to augment acoustic features when training TTS models.
Note that we use the log-Mel spectrogram as the target acoustic features together with continuous log $F_o$~\cite{yu2010continuous}, and voiced/unvoiced flags (V/UV).
Predicting these additional features using the VC model is essential to create emotional TTS models that include $F_o$-dependent high-fidelity neural vocoders~\cite{hwang2021high}.

\begin{figure}[tb]
  \centering
  \includegraphics[width=\linewidth]{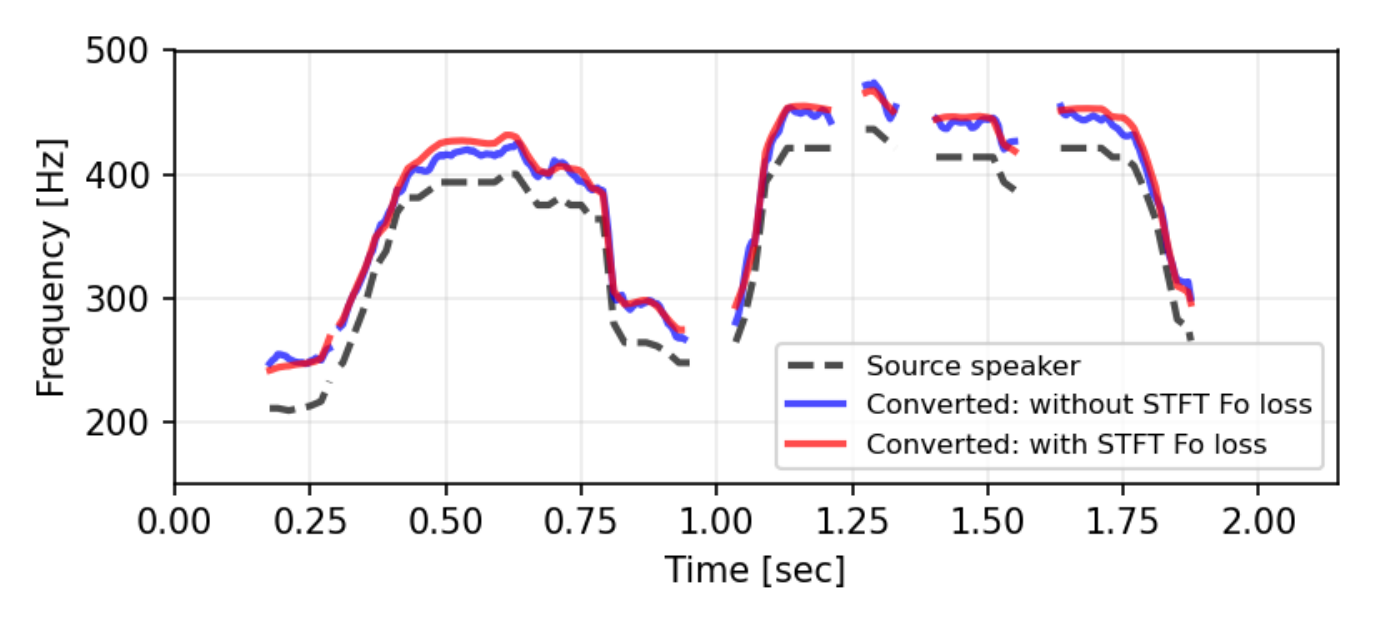}
  \vspace{-8mm}
  \caption{Comparison of the generated $F_o$ with and without the proposed regularization and source $F_o$.}
  \label{fig:f0compare}
  \vspace{-4mm}
\end{figure}

\begin{table*}[t]
\caption{Systems for comparison and the number of utterances for training the VC and TTS models.}
\label{tbl:mos_model}
\begin{center}
\vspace{-15pt}
\scalebox{0.90}{
\begin{tabular}{l|l|cc|ccc|cc|cc} \hline
\multicolumn{1}{c|}{\multirow{3}{*}{Model}} & \multicolumn{1}{c|}{\multirow{3}{*}{Type}} & \multicolumn{2}{c|}{VC training}                  & \multicolumn{7}{c}{TTS training}                                                                                                                                                             \\ \cline{3-11}
\multicolumn{1}{c|}{}                       & \multicolumn{1}{c|}{}                      & \multicolumn{2}{c|}{Neutral}                      & \multicolumn{3}{c|}{Neutral}                                                      & \multicolumn{2}{c|}{Happiness}                       & \multicolumn{2}{c}{Sadness}                         \\ \cline{3-11}
\multicolumn{1}{c|}{}                       & \multicolumn{1}{c|}{}                      & \multicolumn{1}{c}{REC} & \multicolumn{1}{c|}{PS-DA} & \multicolumn{1}{c}{Source} & \multicolumn{1}{c}{Target} & \multicolumn{1}{c|}{VC-DA} & \multicolumn{1}{c}{Source} & \multicolumn{1}{c|}{VC-DA} & \multicolumn{1}{c}{Source} & \multicolumn{1}{c}{VC-DA} \\  \hline
Source	        & Recorded audio	    & -	    & -	     & -	    & -	    & -	    & -	    & -	    & -	    & -	\\ 
SRC-TTS	        & Source speaker TTS	& -	    & -	     & 5.0 K    & -	    & -	    & 2.5 K	& -	    & 2.5 K	& -	\\
TGT-NEU-TTS	    & Target speaker TTS	& -	    & -	     & -	    & 2.5 K	& -	    & -	    & -	    & -	    & -	\\
MS-TTS	        & Multi-speaker TTS 	& -	    & -	     & 5.0 K	& 2.5 K	& -	    & 2.5 K	& -	    & 2.5 K	& -	\\ 
VC-TTS	        & VC-TTS w/o PS         & 2.5 K	& -	     & -	    & 2.5 K	& 5.0 K	& -	    & 2.5 K	& -	    & 2.5 K	\\
VC-TTS-PS	    & VC-TTS w/ PS          & 2.5 K	& 37.5 K & -	    & 2.5 K	& 5.0 K	& -	    & 2.5 K	& -	    & 2.5 K	\\
VC-TTS-PS-1K    & VC-TTS w/ PS          & 1.0 K	& 15.0 K & -	    & 1.0 K	& 5.0 K	& -	    & 2.5 K	& -	    & 2.5 K	\\ \hline

\end{tabular}}	
\scalebox{0.90}{
\begin{threeparttable}
\begin{tablenotes}
\item REC: Recorded data; PS-DA: Data augmented by pitch-shifting; VC-DA: Data augmented by voice conversion
\end{tablenotes}
\end{threeparttable}}
\vspace{-10pt}
\end{center}
\vspace{-10pt}
\end{table*}

\subsubsection{STFT $F_o$ regularization loss function}

To avoid unnatural conversion of the prosody features, we propose an STFT-based $F_o$ regularization loss function.
Following a previous study on a spectrogram domain $F_o$ loss function~\cite{ren2020fastspeech}, we also define the regularization loss function on the spectrogram domain.

Let $X_{n,k}$ and $\hat{X}_{n,k}$ be STFT magnitudes for extracted and predicted $F_o$  sequences for the $n$-th frame index and $k$-th frequency bin, respectively.
The regularization loss is defined as follows:
\begin{equation}
L_\mathrm{F_o} = \frac{1}{M} \sum_{n=1}^{N}\sum_{k=\beta}^{K} \left| \log X_{n,k} - \log \hat{X}_{n,k}\right|,
\label{f0stftloss}
\end{equation}
where $N, K,$ and $M$ represent the number of frames, number of frequency bins, and number of elements in the magnitude, respectively; $\beta$ denotes a hyperparameter that controls the regularization strength. 
To regularize only the fine structure component of $F_o$ (i.e., high-frequency components of the STFT magnitude) that contains little information about speaking styles for reading speech, we set $\beta=3$ based on our preliminary experiments.
Furthermore, we extend the loss function to multiple resolutions inspired by previous studies on multi-level $F_o$ modeling~\cite{ming2015fundamental} and multi-resolution STFT loss~\cite{yamamoto2020parallel}.
Consequently, we optimize the VC model using the proposed regularization loss along with the adversarial loss, cycle consistency loss, and identity mapping loss functions, as described in Scyclone~\cite{kanagaki2020scyclone}.

As shown in Figure~\ref{fig:f0compare}, the $F_o$ trajectory produced without the regularization method oscillates unstably.
By contrast, with the regularization, the stability of the $F_o$ trajectory improves as the VC model can focus on converting essential aspects of prosody variations.

\subsubsection{VC-based data augmentation}

For the criteria described above, we train the Scyclone model using a pair of source and target speaker's speech databases.
Note that the training data consists of neutral recordings and PS-augmented data from each speaker.
As illustrated in Figure \ref{fig:overview}b, we use the resulting VC model to convert all the source speaker's emotional voice into the target speaker's voice.
Simultaneously, to stabilize the training process of the TTS model, we also convert the source speaker's neutral voice to the target speaker's voice.

We use all the converted data, together with the target speaker's neutral recordings, to train the target speaker's emotional TTS system.

\subsection{Text-to-speech}
\label{ssec:tts}

Our TTS model consists of two components: (1) an acoustic model that converts an input phoneme sequence into acoustic features and (2) a vocoder that converts the acoustic feature into the waveform.
For the acoustic model, we use FastSpeech~2~\cite{ren2020fastspeech} with a Conformer encoder~\cite{gulati2020conformer} because of its fast but high-quality TTS capability~\cite{guo2021recent}. 
To adapt FastSpeech~2 for emotional TTS, we condition the model using external emotion code~\cite{choi2019multi}.
For the vocoder, we use the high-fidelity harmonic-plus-noise Parallel WaveGAN (HN-PWG)~\cite{hwang2021high}.

Figure \ref{fig:overview}~(c) shows the training process of TTS with the proposed data augmentation.
We mix synthetic and recorded data for the target speaker and use them to train the acoustic model.
At the inference stage, the TTS model generates emotional speech by inputting text and an emotion code.
Note that we do not use data augmentation for training the vocoder because (1) it has been reported that using a large amount of training data is not crucial for the vocoder~\cite{okamoto2020realtime}, and (2) our preliminary experiments also confirmed subtle improvements when the amount of the source speaker's data was sufficiently large.

\section{Experiments}
\subsection{Experimental setup}
\subsubsection{Database and feature extraction settings}

For the experiments, we used two phonetically and prosodically rich speech corpora recorded by two female Japanese professional speakers, which represent data for the source and target speakers.
We sampled speech signals at 24~kHz with 16~bit quantization. 
The source speaker data contained three speaking styles: \textit{neutral}, \textit{happiness}, and \textit{sadness}, whereas the target speaker data contained only \textit{neutral} style\footnote{
    The average sentence duration for each data set was 5.0, 4.7, 4.2, and 5.1 seconds, respectively.}.

We concatenated the $80$-dimensional log-Mel spectrogram, continuous log $F_o$, and V/UV  with 5~ms analysis intervals as $82$-dimensional features. We used them  as the target acoustic features for both the VC and acoustic models.
We calculated the log-Mel spectrogram with a 40 ms window length.
We extracted $F_o$ and V/UV using the improved time-frequency trajectory excitation vocoder~\cite{song2017effective}.
We obtained $F_o$ for the acoustic features generated by PS data augmentation by shifting the $F_o$ extracted from the original speech. 
We used V/UV extracted from the original speech as the V/UV for the generated data.
We normalized the acoustic features so that they had a zero mean and unit variance for each dimension using the statistics of the training data.

\begin{table*}[t]
\caption{
    Naturalness, speaker similarity, and emotional similarity MOS test results with 95\% confidence intervals.
    Results for the highest score in the VC-based TTS systems are shown in bold.
}
\label{tbl:mos}
\begin{center}
\vspace{-15pt}
\scalebox{0.90}{
\begin{tabular}{l|ccc|ccc|cc}\hline
\multicolumn{1}{c|}{\multirow{2}{*}{Model}} & \multicolumn{3}{c|}{Naturalness}      & \multicolumn{3}{c|}{Speaker similarity}     & \multicolumn{2}{c}{Emotional similarity}                             \\\cline{2-9}
               & {Neutral}            & {Happiness}          & {Sadness}            & {Neutral}           & {Happiness}           & {Sadness}            & {Happiness}          & {Sadness}            \\\hline
{Source}       &         4.88 ± 0.05  &         4.84 ± 0.05  &         4.69 ± 0.07  &         -           &         -             &         -            &         -            &         -            \\
{SRC-TTS}      &         4.56 ± 0.06  &         4.46 ± 0.08  &         4.40 ± 0.08  &         1.15 ± 0.05 &         1.28 ± 0.08   &         1.29 ± 0.08  &         3.48 ± 0.08  &         3.71 ± 0.07  \\
{TGT-NEU-TTS}  &         -            &         -            &         -            &         -           &         -             &         -            &         1.69 ± 0.09  &         1.29 ± 0.07  \\
{MS-TTS}       &         3.91 ± 0.09  &         2.85 ± 0.10  &         2.74 ± 0.11  &         2.85 ± 0.13 &          2.76 ± 0.11  &         2.51 ± 0.12  &         2.48 ± 0.10  &         2.72 ± 0.11  \\
{VC-TTS}       &         4.06 ± 0.08  &         3.88 ± 0.10  & \textbf{4.00 ± 0.10} &         2.98 ± 0.12 &          2.89 ± 0.11  &         3.26 ± 0.11  &         3.33 ± 0.08  &         3.63 ± 0.07  \\
{VC-TTS-PS}    &         4.03 ± 0.08  & \textbf{4.20 ± 0.09} & \textbf{4.00 ± 0.09} &         2.98 ± 0.12 &          2.87 ± 0.12  & \textbf{3.35 ± 0.10} &         3.82 ± 0.05  & \textbf{3.67 ± 0.07} \\
{VC-TTS-PS-1K} & \textbf{4.20 ± 0.08} &         4.08 ± 0.09  &         3.96 ± 0.11  & \textbf{3.19 ± 0.11} & \textbf{2.90 ± 0.12} &         3.31 ± 0.09  & \textbf{3.87 ± 0.04} &         3.63 ± 0.07  \\\hline
\end{tabular}}
\vspace{-20pt}
\end{center}
\end{table*}

\subsubsection{Model details}

For the CycleGAN-based VC model, the generator and discriminator were composed of four and three residual blocks, respectively, and each block consisted of two convolutional layers with leaky ReLU activation. We set the kernel size to three for all the convolutional layers. 
We trained the VC model for 400 K steps using an Adam optimizer~\cite{kingma2014adam}. 
We set the learning rate to 0.0002, and reduced this by a factor of ten every 100 K steps. 
We set the minibatch size to 64.
We set the weight of the proposed regularization loss described in Section \ref{ssec:vc} to 0.1, and used the FFT sizes (32, 64, 128), window sizes (32, 64, 128), and hop sizes (8, 16, 32) for the multi-resolution STFT loss.
We used the identity mapping loss only for the first 10 K steps~\cite{kaneko2019CycleGAN-VC2}.

For the TTS acoustic model, we used four Conformer and Transformer blocks for the encoder and decoder, respectively.
For each block, we set the hidden sizes of the self-attention and feedforward layers to 384 and 1024, respectively. 
To achieve natural prosody for Japanese, for the model, we used accent information as external input~\cite{yasuda2019investigation}.
For emotional TTS, we added emotion embedding followed by a projection layer  with 256-dimensional phoneme and accent embedding.
To improve the duration stability, we used manually annotated phoneme durations.
At the training stage, we used a dynamic batch size with an average of 23 samples to create a minibatch~\cite{hayashi2020espnet}, and trained the models for 200 K steps using the RAdam optimizer~\cite{liu2019radam}.

Table~\ref{tbl:mos_model} summarizes the systems used in our experiments.
We trained the following TTS systems:

\begin{description}
\item[SRC-TTS:] Baseline  TTS model trained with the source speaker's recordings.
\item[TGT-NEU-TTS:] Baseline TTS model trained with the target speaker's recordings (\textit{neutral} style alone).
\item[MS-TTS:] Baseline multi-speaker TTS model trained with source and target speaker's recordings.
\item[VC-TTS:] Baseline TTS model trained with target speaker's recordings and VC-augmented data.
\item[VC-TTS-PS:] Proposed TTS model trained with target speaker's recordings and PS-VC-augmented data
\item[VC-TTS-PS-1K:] Proposed TTS model similarly configured with \textbf{VC-TTS-PS} system, but trained with a limited amount of recordings.
\end{description}

As the vocoder, we trained HN-PWG~\cite{hwang2021high} for 400 K steps with the RAdam optimizer~\cite{liu2019radam}.
For training, we used 5,000 utterances of \textit{neutral} style, 2,500 utterances of \textit{happiness} style, and 2,500 utterances of \textit{sadness} style from the source speaker, and 1,000 utterances of \textit{neutral} style from the target speaker.
We used the same vocoder for all the aforementioned TTS systems.

\subsection{Evaluation}

To evaluate the effectiveness of our proposed method, we conducted subjective listening tests: 5-point naturalness mean opinion score (MOS), 4-point speaker similarity MOS, and 4-point emotional similarity MOS\footnote{
    Following the method used in the VC challenge~\cite{zhao2020voice}, we used the  5-point responses 1 = Bad; 2 = Poor; 3 = Fair; 4 = Good; and 5 = Excellent; and the 4-point responses 1 = Different, absolutely sure; 2 = Different, not sure; 3 = Same, not sure; and 4 = Same, absolutely sure.
    }.

We asked native Japanese raters to make a quality judgment. The number of subjects for each evaluation was 14, 12, and 12, respectively.
For all the tests, we randomly selected 20 utterances from the evaluation set for each system.
In the naturalness evaluation, we evaluated the recorded speech of the source speaker and synthetic speech of five TTS systems, for a total of 360 utterances.
In the speaker similarity evaluation, we set the recorded speech of the target speaker as a reference, and evaluated five TTS systems, for a total of 300 utterances.
Note that the reference samples contained \textit{neutral}, \textit{happiness}, and \textit{sadness} emotions, which constituted of 25 seconds in total.
In the emotional similarity test, we evaluated 240 pairs of utterances that consisted of the recorded emotional speech of the source speaker and the synthetic speech from six TTS systems.
Note that we used the \textit{neutral} TTS system of the target speaker (i.e., \textbf{TGT-NEU-TTS}) as an anchor system only in the emotional similarity test.

\subsection{Results}

The results of the MOS evaluations are shown in Table~\ref{tbl:mos}.
The findings can be summarized as follows: (1) VC data augmentation was effective for improving naturalness and speaker/emotional similarities over the multi-speaker TTS baseline (\textbf{VC-TTS} vs. \textbf{MS-TTS}), particularly for emotional styles; 
and (2) the proposed PS data augmentation further improved performance. In particular, naturalness and emotional similarity significantly improved for \textit{happiness} (\textbf{VC-TTS} vs. \textbf{VC-TTS-PS}) while achieving high emotion reproducibility of the source speaker nearly the same or even better than \textbf{SRC-TTS}.; 
and (3) our proposed method achieved competitive performance, even with a limited number of training data (\textbf{VC-TTS-PS} vs. \textbf{VC-TTS-PS-1K}).
We observed that \textbf{VC-TTS-PS-1K} achieved better naturalness and speaker similarity than \textbf{VC-TTS-PS} for the \textit{neutral} style. 
For naturalness, this could be explained by the source speaker's database having a more natural speaking style than the target speaker, and the style of the source speaker was transferred to the target speaker's TTS when the relative amount of VC augmented data is high (i.e., \textbf{VC-TTS-PS-1K}).
For speaker similarity, we hypothesized that it was caused by the difference of $F_o$ statistics of the training data.
To verify this, we examined the $F_o$ statistics of pseudo \textit{neutral} data used for training \textbf{VC-TTS-PS} and \textbf{VC-TTS-PS-1K}, and found that the latter contained higher $F_o$ on average 4.04~Hz. 
Because $F_o$ of the target speaker was higher than that of the source speaker in our experiments, higher-pitched samples of \textbf{VC-TTS-PS-1K} tended to have higher speaker similarity for the \textit{neutral} style.
We encourage readers to listen to the samples provided on our demo page\footnote{\url{https://ryojerky.github.io/demo_vc-tts-ps/}}. 

\begin{figure}[!t]
   	\begin{minipage}[b]{1.0\linewidth}
        \centering
   	    \includegraphics[width=0.9\linewidth]{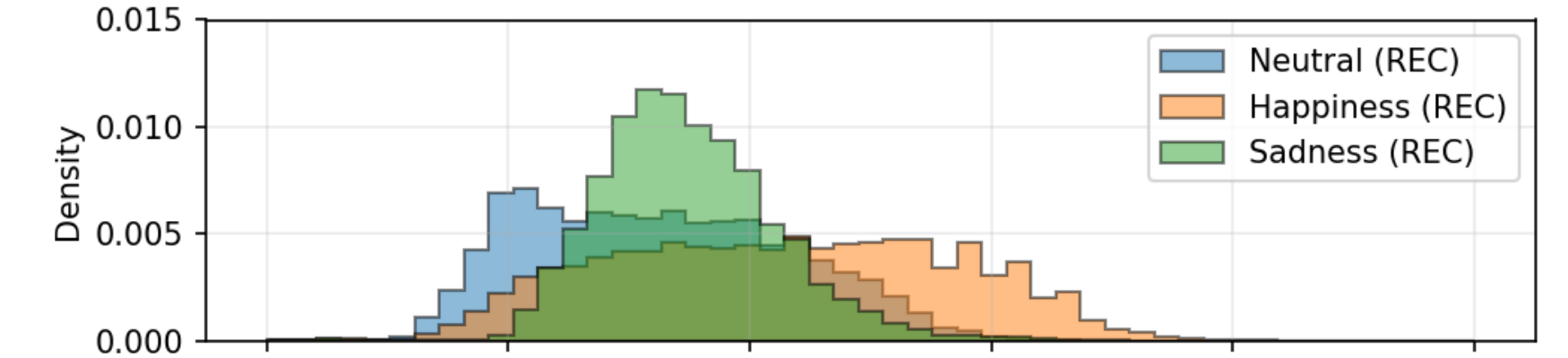}
   		\centerline{(a)}  \medskip
        \vspace{-5mm}
    \end{minipage}
   	\begin{minipage}[b]{1.0\linewidth}
        \centering
   	    \includegraphics[width=0.9\linewidth]{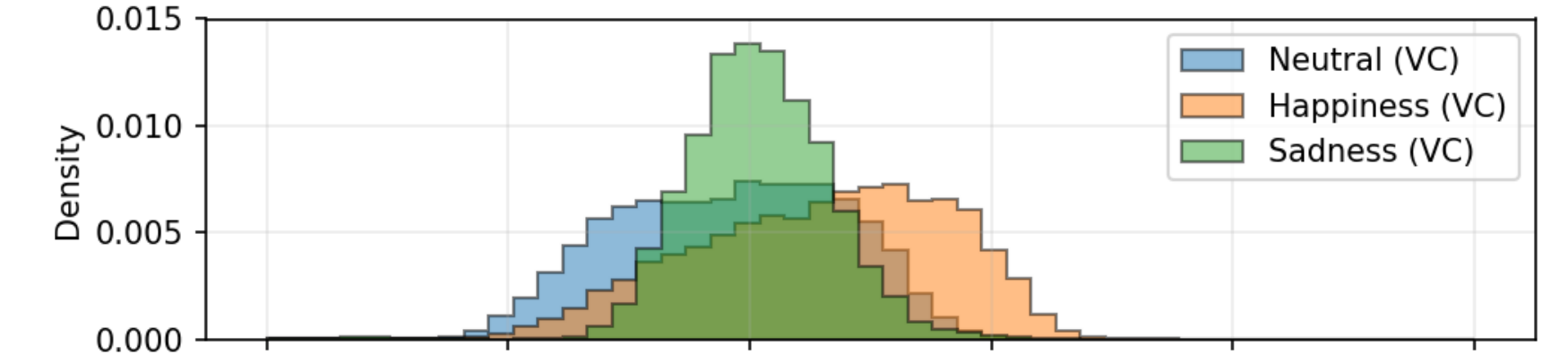}
   		\centerline{(b)}  \medskip
        \vspace{-5mm}
   	\end{minipage}
   	\begin{minipage}[b]{1.0\linewidth}
        \centering
   	    \includegraphics[width=0.9\linewidth]{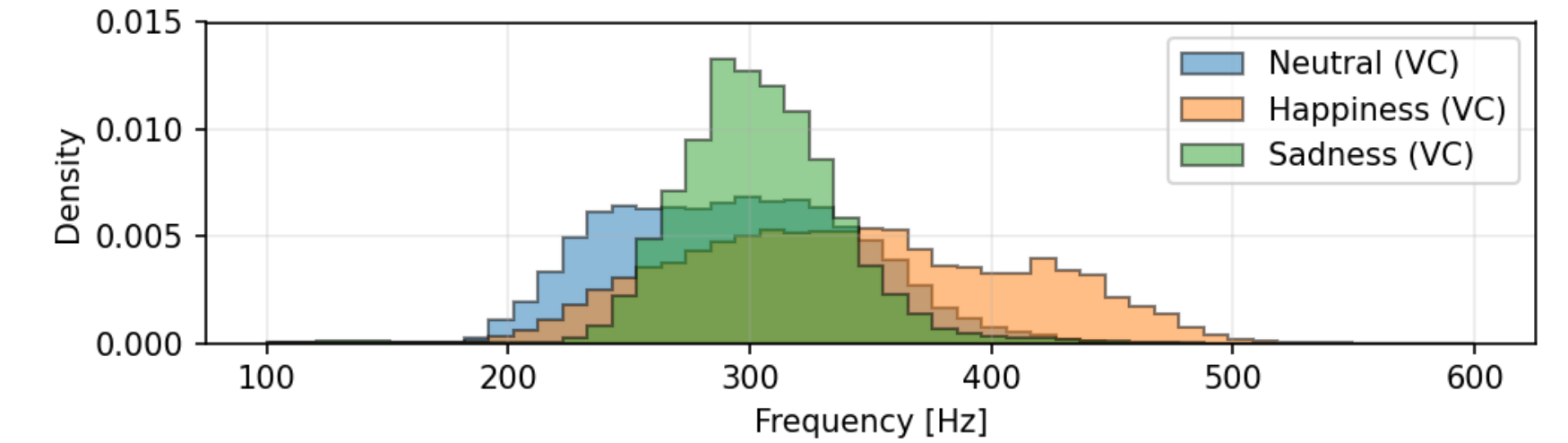}
   		\centerline{(c)}  \medskip
        \vspace{-2mm}
   	\end{minipage}
    \vspace{-10mm}
    \caption{$F_o$ distributions obtained from each emotion: (a) the source speaker's recorded data, (b) the target speaker's VC-augmented data, and (c) that with the proposed PS method.}
    \label{fig:f0-distribution}
    \vspace{-4mm}
\end{figure}

To further verify the effectiveness of the proposed method, we analyzed the $F_o$ distributions of the original data and the pseudo data generated by the VC model.
As illustrated in  Figure~\ref{fig:f0-distribution}, the distribution of $F_o$ with PS data augmentation was closer in shape to that of the original data for \textit{happiness}. The results confirmed that the VC model trained on the proposed PS-augmented data generated richer pitch variations that were close to natural recordings compared with the VC model trained without PS data augmentation.
By contrast, we observed similar distributions with and without the proposed PS augmentation for \textit{sadness}. 
This can be explained as follows: \textit{sadness} is less dynamic and has fewer pitch variations than \textit{happiness}.
The results suggest that our proposed method was particularly suited for emotionally expressive and dynamic styles, such as \textit{happiness}.

\section{Conclusion}

We proposed a cross-speaker emotion style transfer method for a low-resource expressive TTS system, where expressive data is not available for the target speaker.
Our proposed method combines PS-based and VC-based augmentation methods to stabilize training for both VC and TTS acoustic models.
Subjective test results showed that the FastSpeech~2-based emotional TTS system learned by the proposed method improved naturalness and emotional similarity compared with conventional methods.
In the future, we aim to apply the proposed method to more distinctive, expressive, and dynamic styles of speech.

\section{Acknowledgments}
This work was supported by Clova Voice, NAVER Corp., Seongnam, Korea. 

\section{References}
{
\setstretch{1.0}
\printbibliography
}

\end{document}